\documentclass[aps,prb,twocolumn,showpacs,superscriptaddress]{revtex4}

\usepackage{graphicx}
\usepackage{dcolumn}

\usepackage{eucal}
\usepackage[dvips]{epsfig}
\usepackage{changebar}
\usepackage{amssymb}

\usepackage{amsmath}
\newcommand{\mbf}[1]{\mathbf{#1}}

\newcommand{\EW}[1]{\left\langle{#1}\right\rangle}

\newfont{\tensy}{cmsy10}
\newcommand{\chemical}[1]{$\fontdimen16\tensy=3.0pt
                       \fontdimen17\tensy=3.0pt\mathrm{#1}$}

\newcommand{\kB}{k_B}
\newcommand{\Tc}{T_C}
\newcommand{\cP}{c_P}
\newcommand{\GaMnAs}{\chemical{Ga_{1-{\it x}}Mn_{\it x}As}}
\newcommand{\GaMnN}{\chemical{Ga_{1-{\it x}}Mn_{\it x}N}}

\begin{document}

\title{Magnetism in (III, Mn)-V Diluted Magnetic Semiconductors: Effective Heisenberg Model}

\author{S. Hilbert}
 \email{hilbert@mpa-garching.mpg.de}
\affiliation{Max-Planck-Institut f{\"u}r Astrophysik, 
 Karl-Schwarzschild-Stra{\ss}e 1, D-85741, Garching, Germany}
\author{W. Nolting}
\affiliation{Institut f\"ur Physik, Humboldt-Universit\"at zu Berlin,
 Newtonstra{\ss}e 15, D-12489 Berlin, Germany}
\date{\today}

\begin{abstract}
The magnetic properties of the diluted magnetic semiconductors (DMS)
(Ga, Mn)As and (Ga, Mn)N are investigated by means of an
effective Heisenberg model, whose exchange parameters are obtained from
first-principle calculations. The finite-temperature properties of the
model are studied numerically using a method
based upon the Tyablikov approximation. The method properly
incorporates the effects of positional disorder present 
in DMS.
The resulting Curie temperatures for (Ga, Mn)As are in excellent
agreement with experimental data. Due to percolation effects and
noncollinear magnetic structures at higher Mn concentrations, our calculations
predict for (Ga, Mn)N very low Curie temperatures compared to mean-field
estimates. 
\end{abstract}


\maketitle

%%%%%%%%%%%%%%%%%%%%%%%%%%%%%%%%%%%%%%%%%%%%%%%%%%%%%%%%%%%%%%%%%%%%%%

\section{Introduction}
\label{sec:Introduction}
Ferromagnetic (III, Mn)-V diluted magnetic semiconductors (DMS) have
attracted considerable attention among scientist during
the past years \cite{Ohno99,Dietl02}. Their investigation has been
driven by the idea of using their coupled electronic and magnetic
degrees of freedom to construct  electronic devices ranging from fast
nonvolatile memories to quantum computers \cite{ZuticEtal04}. 
To date, however, technical applicability has been limited by the 
fact that most known DMS have Curie temperatures $\Tc$ below room
temperature\cite{ReedEtal01,Dietl02,EdmondsEtal02,KuEtal03,ParkEtal02}. 

For the development of ferromagnetic DMS with higher Curie
temperatures, it is important to understand theoretically the
magnetism in these materials and to develop theories which provide
reliable qualitative {\em and} quantitative predictions.
The magnetism in these materials is due to magnetic moments localized
at magnetic impurities, which interact with each other indirectly
via holes in the valence and impurity band of the host
semiconductor. Therefore, for the description, one often employs an
effective Heisenberg model, whose exchange parameters are determined
by the interaction between the localized moments 
and the holes \cite{DietlEtal00,KudrnovskyEtal04,BouzerarPareek02,
  LitvinovDugaev01,KaminskiEtal04,SchliemannEtal01,BreyGomezsantos03,
  BergqvistEtal04}. 
However, the magnetic impurities are mainly randomly distributed
over the sites of the crystal lattice. This positional disorder breaks
the translational symmetry of the crystal and thus greatly complicates
the theoretical description of the material.
Studies based on the mean-field approximation (MFA)
\cite{DietlEtal00,KudrnovskyEtal04} or the random-phase approximation
(RPA) combined with the virtual-crystal approximation (VCA)
\cite{BouzerarPareek02} neglect effects of the positional disorder in
DMS. Approaches based on percolation theory
\cite{LitvinovDugaev01,KaminskiEtal04} account for the randomness of
the impurity positions, but require a simple functional dependence of
the exchange parameters on the inter-spin distance and treat the
magnetism itself only on a mean-field level. Monte-Carlo simulations (MC) 
\cite{SchliemannEtal01,BreyGomezsantos03,BergqvistEtal04,SatoEtal04}
seem to provide a better way to include the positional disorder, but
these are numerically expensive and usually assume classical
spins. However, a proper treatment
of the positional disorder of the localized moments and their quantum
nature is needed to make reliable predictions about the magnetic
properties of DMS \cite{Schliemann03,Timm03}.

In a previously published article \cite{KudrnovskyEtal04}, the
exchange parameters of an effective (classical) Heisenberg Hamiltonian
have been calculated from first-principles for \GaMnAs{} and
\GaMnN{}. There, however, these had only been used to calculate Curie
temperatures within MFA. More recently, results of classical MC
simulations on the basis of these exchange parameters have been
presented \cite{BergqvistEtal04}. Here, we employ a different approach
\cite{Hilbert03,HiNo04} to investigate the properties of the effective
Heisenberg Hamiltonian. This approach generalizes the Tyablikov
approximation \cite{BogolyubovTyablikov59} to systems with positional
disorder, which is treated numerically exactly. Furthermore, 
the method assumes quantum spins. The quantum fluctuations of the
spins are treated within random-phase 
approximation, which goes beyond MFA and the classical-spin
approximation. It should be mentioned that a
similar approach has been proposed in \cite{BouzerarZimanKudrnovsky04}.

\section{Model}
\label{sec:Model}
Details of the electronic-structure calculation for \GaMnAs{} and
\GaMnN{} and the extraction of the exchange parameters
\mbox{$J(\mbf{R})$} as function of the Mn-Mn distance $\mbf{R}$ can be
found in Ref. \cite{KudrnovskyEtal04}. 
Here, these exchange parameters are used as input for a `diluted' Heisenberg model 
\begin{equation}
\label{eq:Heisenberg_Hamiltonian}
  H=-\sum_{i,j=1}^{N} J_{ij}\, \mbf{e}_i\cdot \mbf{e}_j\;,
\end{equation}
in which only a fraction of the lattice sites is occupied by a spin.
Hence, $i$ and $j$ label the occupied lattice sites only, whose total
number is $N$, and \mbox{$\mbf{e}_i=\left(S_i^x,S_i^y,S_i^z\right)/(\hbar S)$} 
is the normalized spin operator of the localized magnetic moment at lattice site
$i$ with lattice vector $\mbf{R}_i$ and
\mbox{$J_{ij}=J(|\mbf{R_i}-\mbf{R}_j|)$}. The magnitude $S$ of the
spins is absorbed by the exchange parameters due to the particular way
in which these are calculated from the electronic structure.

The finite-temperature properties of Hamiltonian
\eqref{eq:Heisenberg_Hamiltonian} are studied using a generalization
of the Tyablikov approximation to systems without translational
symmetry \cite{Hilbert03,HiNo04}. 
The generalization treats the
positional disorder in the spin system numerically exactly except that a
uniform magnetization is assumed. Furthermore, 
the effects of low-energy quantum excitations, i.e. magnons, are
included. Within this approximation, the local magnon spectral density 
is given by \cite{Hilbert03,HiNo04}:
\begin{equation}
S_{ii}\left(E\right)= 2\hbar^2\EW{S^z}\frac{1}{N}\sum_{r=1}^{N}
  \delta\left(E- \frac{2\hbar\EW{S^z}}{\hbar^2 S^2}E_r\right)\;,
\end{equation}
where the $E_r$ are the eigenvalues of the
Hamilton matrix $\mbf{H}$, which is defined by its matrix elements
\mbox{$H_{ij}=\delta_{ij}\sum_{n=1}^{N}J_{in}-J_{ij}$}.
These eigenvalues also determine the Curie temperature: 
\begin{equation}
  \label{eq:Tc}
  {\kB} \Tc=\frac{2}{3}\frac{S\left(S+1\right)}{S^2}
  \left(\frac{1}{N}\sum_r\frac{1}{E_r} \right)^{-1}\;.
\end{equation}
To evaluate this expression for a given set of $E_r$'s, the value of
$S$ has to be fixed. 
For  Mn ions in \GaMnAs{} and \GaMnN{}, $S=5/2$ should be
appropriate \cite{Dietl02}. However, this choice is not consistent
with the calculation of the exchange parameters from the electronic
structure, where classical spins are assumed.
Therefore, we will use Eq. \eqref{eq:Tc} in the limit
$S\to\infty$, which yields $\Tc$ values a factor $5/7$ less than
for $S=5/2$.
 
Due to the positional disorder of the spins present in DMS,
the eigenvalues cannot be computed by Fourier transformation of
$\mbf{H}$.  
However, the eigenvalues may be obtained by the numerical
diagonalization of the Hamilton matrix for a finite system. In our
calculations, we used systems of $\sim10\,000$ spins, which were randomly
distributed over the lattice sites of a cubic section of an fcc lattice
with periodic boundary conditions. For each concentration $x$ of
Mn ions, we averaged the spectral densities over eight random
configurations.

\section{Results}
\label{sec:Results}

\begin{figure}[tb]
\centerline{\includegraphics[width=1\linewidth]
{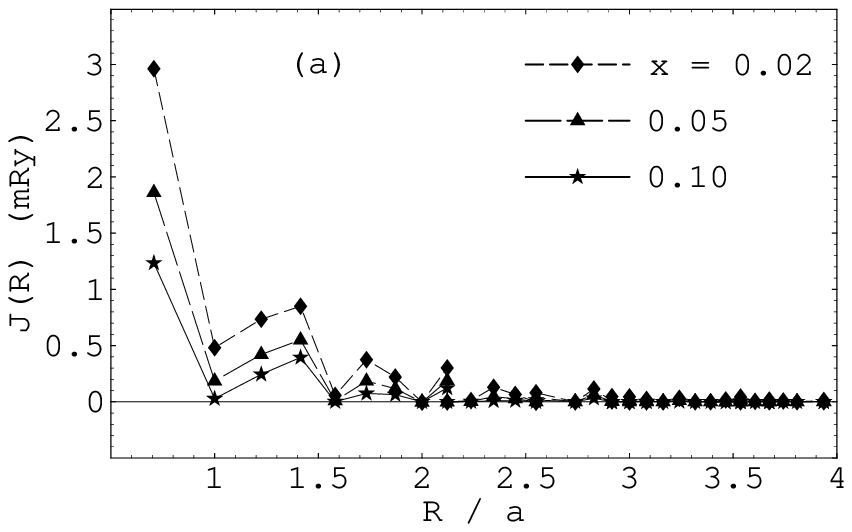}}
\vspace{1ex}
\centerline{\includegraphics[width=1\linewidth]
{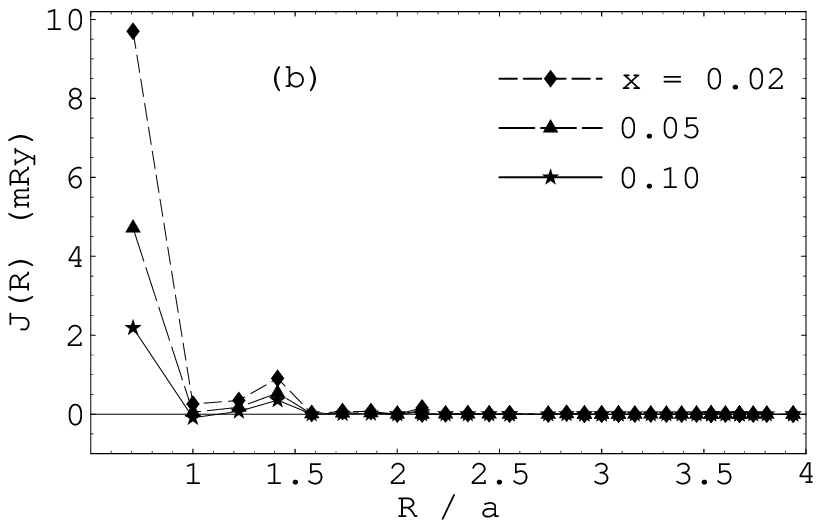}}
\caption{
\label{fig:J}
Exchange interactions $J(\mbf{R})$ between Mn ions of distance
$\mbf{R}$  in (a) \GaMnAs{} and (b) \GaMnN{}  for various concentrations $x$ 
(from \cite{KudrnovskyEtal04,KudrnovskyPrivate04})} 
\end{figure}

In Fig. \ref{fig:J}, the Mn-Mn exchange interactions $J(\mbf{R})$ in
\GaMnAs{} and in \GaMnN{} are shown as functions of the Mn-Mn distance $R$ for
several concentration $x$. In \GaMnAs{}, the
falloff of the interaction with $R$ is comparably slow.
In \GaMnN{}, the interaction between nearest neighbors is much larger than in
\GaMnAs{}, but Mn moments further apart are only very weakly coupled. 

\begin{figure}[tb]
\centerline{\includegraphics[width=1\linewidth]
{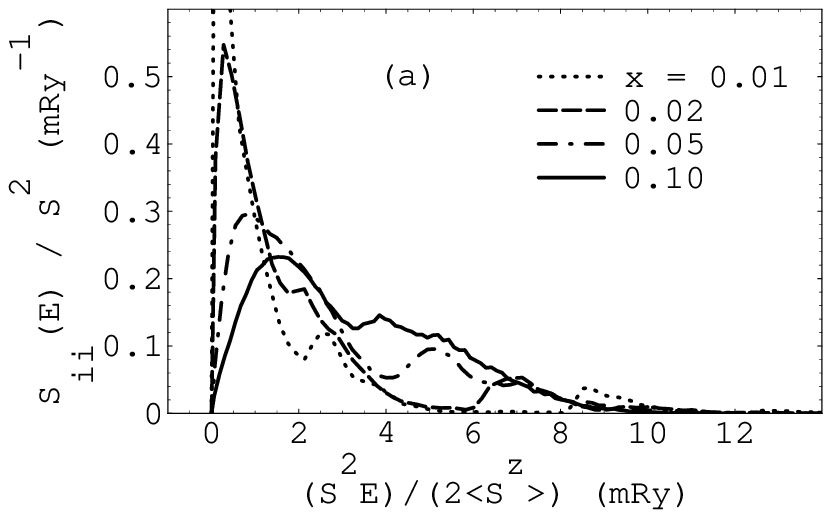}}
\vspace{1ex}
\centerline{\includegraphics[width=1\linewidth]
{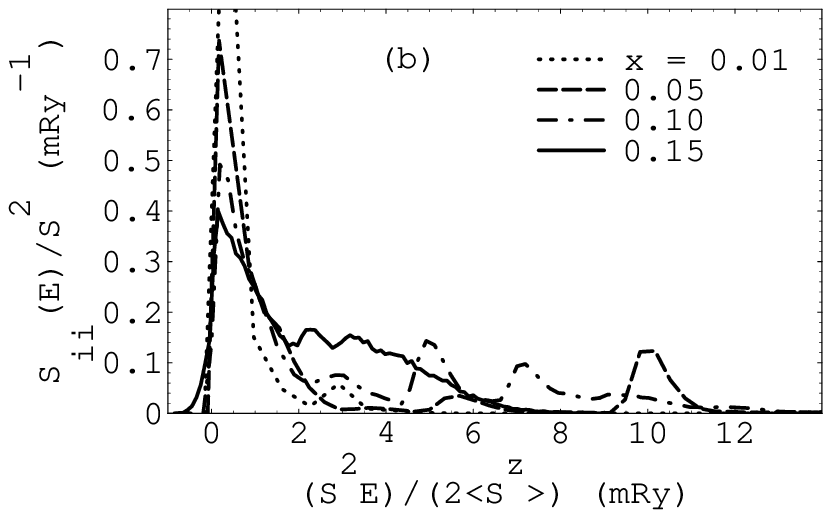}}
\caption{
\label{fig:Sii}
  Local magnon spectral density $S_{ii}(E)$
for (a) \GaMnAs{} and (b) \GaMnN{} for various concentrations $x$ of
Mn
} 
\end{figure}

Figure \ref{fig:Sii} shows the resulting magnon spectral densities. For
\GaMnAs{}, the spectrum is smooth and continuous. 
For \GaMnN{}, one can recognize remnants of peaks typical for
nearest-neighbor interaction at low concentrations, which are
broadened by small long-ranged interactions. Compared to \GaMnAs,
there is a large spectral density at low energies for \GaMnN{}. For
concentrations $x\geq0.08$, antiferromagnetic interactions come into play 
and negative magnon energies appear indicating a ground state which is
not a saturated ferromagnet \cite{HiNo04}.

\begin{figure}[tb]
\centerline{\includegraphics[width=1\linewidth]
{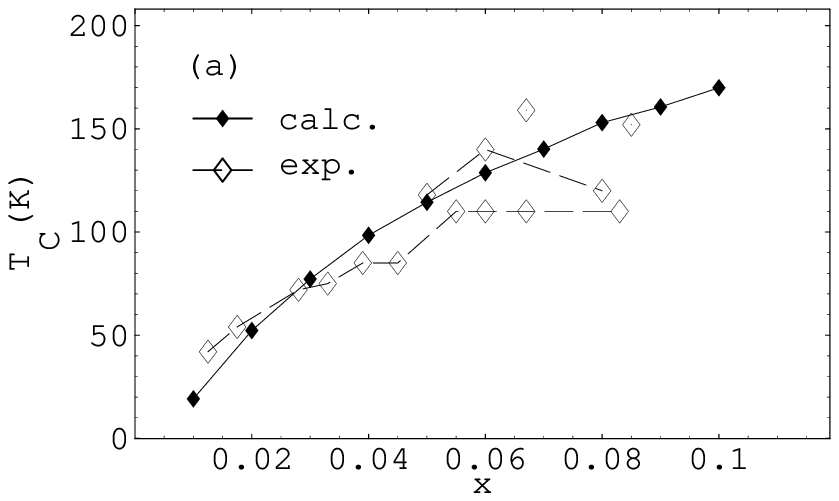}}
\vspace{1ex}
\centerline{\includegraphics[width=1\linewidth]
{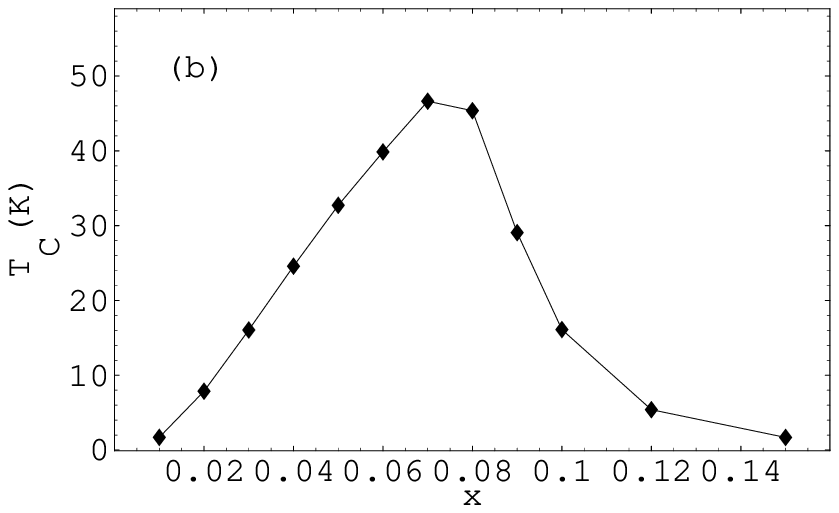}}
\caption{
\label{fig:Tc}
  Calculated Curie temperature $\Tc$ of
  (a) \GaMnAs{} (compared with experimental values of annealed samples
  \cite{PotashnikEtal02,EdmondsEtal02,EdmondsEtal04,KuEtal03})
  and (b) \GaMnN{} for various concentrations $x$ of Mn ions 
} 
\end{figure}

The Curie temperatures calculated using  Eq. \eqref{eq:Tc} are shown in
Fig. \ref{fig:Tc}. For \GaMnAs{}, the calculated values agree remarkably
well with the experimental values of optimally annealed samples
\cite{PotashnikEtal02,EdmondsEtal02,KuEtal03}. Furthermore, the calculated
curve suggests that slightly higher $\Tc$'s might be achieved by further
increasing the Mn content $x$, but values above 300K seem rather
unlikely.

Since experimental values for $\Tc$ in \GaMnN{} are quite controversial
(reported values range from 0K to 940K 
\cite{TheodoropoulouEtal01,OverbergEtal01,ReedEtal01,SonodaEtal02,ThalerEtal02,PloogEtal03}),
we refrain from a comparison here. 
However, the Curie temperatures we calculated are quite low
compared to earlier mean-field estimates (e.g. in
Ref. \cite{Dietl02}).
These low $\Tc$ values despite the high
values of the nearest-neighbor exchange may be explained as follows:
For concentrations well below the nearest-neighbor percolation
threshold $\cP\approx0.2$ \cite{Stauffer95}, even a large nearest-neighbor exchange
does not contribute substantially to the stability of the magnetic
phase. Since the exchange parameters for larger inter-spin distances
are very small in \GaMnN{}, ferromagnetic order can only be established
at very low temperatures.
Note that the drop of $\Tc$ for $x\geq0.08$ may be due to the used
approximation: As indicated by the magnon spectra 
seen in Fig. \ref{fig:Sii}, the system's ground state is different
from a saturated ferromagnet, but a such uniform magnetic state is
assumed in the approximation.

\begin{figure}[tb]
\centerline{\includegraphics[width=1\linewidth]
{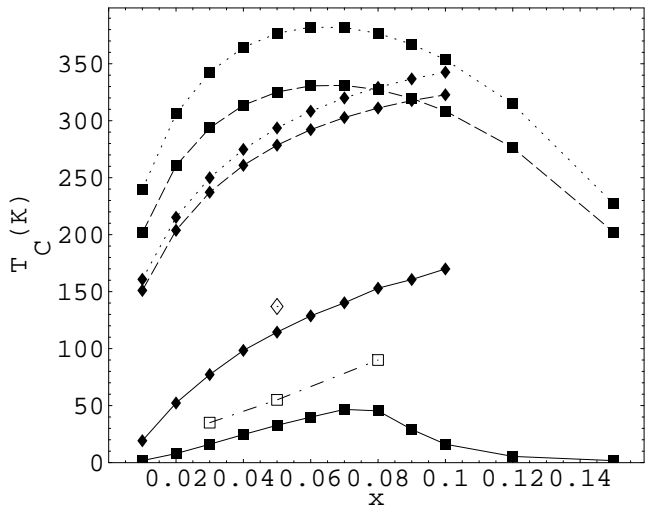}}
\caption{
\label{fig:Tc_compared}
 Comparison of the Curie temperatures $\Tc$ of \GaMnAs{} (diamonds) and
 \GaMnN{} (squares) obtained
 by the presented approach (solid line, filled symbols), VCA-RPA
 (dashed line, filled symbols),  MFA  (dotted line, filled symbols)
 and MC (dash-dotted line, open symbols, taken
 from \cite{BergqvistEtal04}).
} 
\end{figure}

Figure \ref{fig:Tc_compared} presents a comparison of the 
the Curie temperatures calculated using different approximations for
the effective Heisenberg model. The $\Tc$ values obtained by MC simulations
are slightly higher than the ones calculated by the presented approach,
whereas both MFA and VCA-RPA yield much higher $\Tc$'s.
For \GaMnAs{}, the difference is about a factor two to eight. 
For \GaMnN{}, the difference is even 
much larger. This is due to the fact that the MFA and VCA-RPA do not take into
account percolation effects: Large nearest-neighbor interactions yield
large Curie temperatures even for concentrations well below the
nearest-neighbor percolation threshold. However, for such
concentrations, the nearest-neighbor interaction strength should not
play an important role for the ferromagnetic stability, which can be
easily seen by considering the case of nearest-neighbor interaction
only \cite{HiNo04}.

\section{Summary}
\label{sec:Summary}

In this paper, we presented a method for calculating the magnetic
properties of ferromagnetic DMS. The method applies a Tyablikov-like
approximation for systems with positional disorder to an effective
Heisenberg Hamiltonian, whose exchange
parameters where obtained by first-principle calculations.
Unlike in MFA or VCA-RPA, no approximations with respect to the
positional disorder are made apart from the simplification of a uniform 
magnetization. As the main advantage over classical MC simulations, the
presented treatment of the effective Heisenberg model admits quantum
spins and thus may open up a way towards a fully quantum-mechanical
treatment of magnetism in DMS. 
Furthermore, the numerical effort is fairly low compared to MC
simulations. 

Our calculations of $\Tc$ for \GaMnAs{} show excellent agreement with
experimental data. For \GaMnN{}, we obtained very low Curie
Temperatures despite high effective nearest-neighbor exchange
parameters, which shows the importance of percolation effects. Moreover,
for both \GaMnAs{} and \GaMnN{}, the $\Tc$ values we found are much lower
than MFA and VCA-RPA values. These results support recent findings
obtained by using MC simulations in combination with first-principle
methods \cite{BergqvistEtal04,SatoEtal04}.

The presented model should be improved by using a self-consistent method
describing the electronic degrees of freedom at finite
temperature (such as, e.g., in
\cite{NoltingEtal97,SantosNolting02}). In order to obtain
a fully quantum mechanical theory, quantum spins 
should be used instead of classical spins in 
in the calculation of the effective exchange parameters from the
electronic structure. This will also remove the ambiguity in the choice
of $S$. 
Furthermore, the treatment of the effective Heisenberg model may be
extended to allow for a site-dependent $\EW{S_i^z}$. 
In addition, the model might be improved in order
to handle systems with a ground state deviating from a saturated ferromagnet.
Furthermore, clustering and
other forms of short-range chemical ordering may also be included into
the model in order to investigate their effects on the magnetic stability.
Finally, the method should be applied to other DMS.

%%%%%%%%%%%%%%%%%%%%%%%%%%%%%%%%%%%%%%%%%%%%%%%%%%%%%%%%%%%%%%%%%%%%%%

\begin{acknowledgments}
This work benefited from the support of the SFB290 of the Deutsche
Forschungsgemeinschaft. Thanks to J. Kudrnovsk{\'y} for
supplying the exchange parameters and for helpful discussion.
\end{acknowledgments}

%%%%%%%%%%%%%%%%%%%%%%%%%%%%%%%%%%%%%%%%%%%%%%%%%%%%%%%%%%%%%%%%%%%%%%%%%%%

\end{document}